\documentclass[aps,prb,twocolumn,10pt,superscriptaddress,floatfix,nofootinbib]{revtex4-2}

\usepackage{amsmath,amssymb,bm}
\usepackage{graphicx}
\usepackage{tikz}
\usetikzlibrary{arrows.meta}
\colorlet{fci}{gray!18}
\colorlet{sc}{blue!14}
\usepackage{hyperref}
\hypersetup{hidelinks}

\begin{document}

\title{Parafermions in fractional Chern insulator--superconductor
heterostructures: the role of spin polarization}

\author{Aaron Amire}
\email{realaaronamire@gmail.com}
\affiliation{Independent researcher, Shanghai, China}

\date{\today}

\begin{abstract}
Most proposals for $\mathbb{Z}_3$ parafermions in fractional quantum
Hall--superconductor structures used the spin-unpolarized $\nu=2/3$ Halperin
$(1,1,2)$ state. The fractional quantum anomalous Hall (FQAH) states of twisted
MoTe$_2$ and rhombohedral graphene are believed to be spin- and
valley-polarized Jain states, with the same topological order but a different spin
structure. We analyze a trench between two polarized $\nu=2/3$ edges. Among the
pairing and tunneling bilinears of the edge-mode electrons, exactly two of each have
zero conformal spin. Pairing is compatible only with the neutral-mode
superconductivity channel and tunneling only with neutral-mode backscattering,
whereas in the unpolarized state singlet pairing and tunneling select the same
channel; the $(1,1,2)$ singlet pair is a higher-order dressed operator on the
polarized trench. Since the two channels condense the same neutral anyon up to a local
operator, the domain walls carry $\mathbb{Z}_3$ parafermions, 3 protected states
per pair, whether or not the neutral channel changes between regions, and
fixing the fermion parity does not enlarge this count. Before the neutral gap
forms, pairing is at best marginal for repulsive inter-edge coupling unless an
attractive neutral coupling favors its channel; once the gap has formed, both charge terms are relevant for an inter-edge
Coulomb coupling below $7/25$ of the intra-edge one. The
superconductor must supply even-frequency, equal-spin, intravalley pairing, which an
$s$-wave singlet superconductor with only Ising spin--orbit coupling does not
provide at leading order, whereas the chiral $p+ip$ and $f+if$ states proposed for
the adjacent superconductor do. The $\mathbb{Z}_3$ label is carried by a
quasiparticle of the FQAH state, so no process confined to a junction whose tunneling segments are not FQAH can change it, and we expect a $6\pi$-periodic Josephson current with or
without fixed fermion parity.
\end{abstract}

\maketitle

\section{Introduction}

The fractional quantum anomalous Hall (FQAH) effect has been observed at zero
magnetic field in twisted bilayer MoTe$_2$~\cite{Cai2023,Zeng2023,Park2023,Xu2023}
and in rhombohedral multilayer graphene aligned with hBN~\cite{Lu2024}, which makes
fractional Chern insulators (FCIs) an experimental platform. Independently, it has
been shown that coupling fractional quantum Hall (FQH) edges to a superconductor
produces parafermion zero modes at domain walls between pairing- and
backscattering-dominated regions~\cite{Lindner2012,Clarke2013,Vaezi2013}, with
$\mathbb{Z}_3$ parafermions at $\nu=2/3$~\cite{Vaezi2014,Mong2014,AliceaFendley2016}.
Networks of these modes can host Fibonacci anyons, which are universal for
topological quantum computation by braiding alone~\cite{Mong2014}.

Whether FCIs can serve as the host material for these proposals depends on whether
the FQH construction applies to them. It was formulated for the spin-unpolarized
$(1,1,2)$ state~\cite{Mong2014,AliceaFendley2016}, whose edges carry electrons of
both spins, so that an $s$-wave singlet superconductor can pair electrons across a
narrow trench. The FQAH states, on the other hand, are spontaneously spin- and valley-polarized~\cite{Cai2023,Zeng2023,Park2023,Xu2023,Lu2024}
and are believed to belong to the Jain sequence~\cite{Bollmann2026}, which changes
the set of allowed pairing terms.

Quantitative treatments of the
construction~\cite{Lindner2012,Clarke2013,Vaezi2014,Mong2014,Alavirad2017,Repellin2018,Cao2024,Yoshitome2025,Bollmann2026}
have used a spin-unpolarized parent state, a Laughlin $\nu=1/m$ state without a
neutral mode, the boundary between spin-polarized and spin-unpolarized $\nu=2/3$
phases, where the counter-propagating modes carry opposite spins~\cite{Liang2019},
or a polarized $\nu=2/3$ state whose neutral mode is assumed to be gapped by an
unspecified mechanism~\cite{Bollmann2026}. The gapping of the neutral modes alone
was studied in Refs.~\cite{Vayrynen2022,Park2024}. We are not aware of an analysis
of how pairing and tunneling on a polarized $\nu=2/3$ Jain trench select and
generate the neutral gap, of the zero modes that result, or of the pairing symmetry
the superconductor has to provide, and we address these questions here.

Section~\ref{sec:construction} reviews the $\nu=2/3$ edge theory and the
parafermion construction, and Sec.~\ref{sec:transfers} the ingredients that carry
over to FCIs. Section~\ref{sec:fails} contains the main calculation: the local
operators on a polarized trench, the neutral channels they select and generate, the
resulting zero modes, the effect of inter-edge Coulomb coupling, and the conditions
on the superconductor. Rough energy scales are estimated in Sec.~\ref{sec:energy}. Experimental signatures, including the protection of the fractional Josephson period against processes confined to the junction, are discussed in
Sec.~\ref{sec:signatures}, and Sec.~\ref{sec:alternative} compares with intrinsic
non-Abelian FCIs.

\section{The \texorpdfstring{$\nu=2/3$}{nu=2/3} edge and the parafermion construction}
\label{sec:construction}

\subsection{Edge theory}

An Abelian quantum Hall state is characterized by an integer symmetric $K$-matrix
and a charge vector $\bm{t}$, with $\nu=\bm{t}^{\mathsf{T}}K^{-1}\bm{t}$, and its
edge is described by the Lagrangian
\begin{equation}
\mathcal{L}=\frac{1}{4\pi}K_{IJ}\,\partial_t\phi_I\partial_x\phi_J
-\frac{1}{4\pi}V_{IJ}\,\partial_x\phi_I\partial_x\phi_J .
\label{eq:edgeL}
\end{equation}
The signature of $K$ determines the chirality of the modes, and the parity of its
diagonal entries distinguishes fermionic (odd) from bosonic (even) states.

For the hole-conjugate $\nu=2/3$ state,
\begin{equation}
K=\mathrm{diag}(1,-3),\qquad \bm{t}=(1,1),
\label{eq:Kcorrect}
\end{equation}
which gives $\nu=1-\tfrac13=\tfrac23$. The diagonal entries are odd and the
signature is $(+,-)$, so the two modes counter-propagate. An equivalent form is
$K'=\bigl(\begin{smallmatrix}1&2\\2&1\end{smallmatrix}\bigr)$, obtained from
Eq.~\eqref{eq:Kcorrect} as $K'=WKW^{\mathsf{T}}$, $\bm{t}'=W\bm{t}$ with
$W=\bigl(\begin{smallmatrix}1&0\\2&-1\end{smallmatrix}\bigr)\in GL(2,\mathbb{Z})$,
$\det W=-1$. This $W$ leaves $\bm{t}=(1,1)$ unchanged, while
$\bigl(\begin{smallmatrix}1&0\\2&1\end{smallmatrix}\bigr)$ also yields $K'$ but
maps $\bm{t}$ to $(1,3)$.

The matrix $K'$ also describes the spin-unpolarized Halperin $(1,1,2)$ state, whose
two components are spin-up and spin-down electrons. Its edge carries a charge mode
$\phi_\uparrow+\phi_\downarrow$ and a counter-propagating neutral mode
$\phi_\uparrow-\phi_\downarrow$~\cite{Mong2014}. Up to $GL(2,\mathbb{Z})$, the
polarized and unpolarized $\nu=2/3$ states thus have the same $K$-matrix and charge
vector and differ only in the spin carried by the modes, which $K$ does not encode. The edge excitations of the two states have been compared microscopically in Ref.~\cite{WuSreejithJain2012}.
In the unpolarized state the neutral mode is a spin mode, while in the polarized
state both modes lie in a single spin and valley sector. The consequences are
worked out in Sec.~\ref{sec:fails}. The parafermion construction was formulated for
the $(1,1,2)$ state~\cite{Mong2014,AliceaFendley2016}; see also
Ref.~\cite{Vaezi2014}.

Random tunneling between the two modes of a real edge drives the $\nu=2/3$ edge to
the disorder-dominated fixed point of Kane, Fisher and Polchinski
(KFP)~\cite{KFP1994}, at which the modes reorganize into a downstream charge mode
with conductance $\tfrac23 e^2/h$ and an upstream neutral mode. The upstream
neutral mode has been detected experimentally~\cite{Sabo2017,Bhattacharyya2019}.
The KFP fixed point equilibrates this mode but does not gap it
(Sec.~\ref{sec:neutral}).

\subsection{Two-edge geometry}
\label{sec:geometry}

\begin{figure}[t]
\centering
\resizebox{\columnwidth}{!}{%
\begin{tikzpicture}[x=1cm,y=1cm,font=\footnotesize,
  chg/.style={thick,-{Latex[length=1.6mm]}},
  neu/.style={thick,dashed,-{Latex[length=1.6mm]}}]
\node[anchor=west] at (-0.35,2.75) {(a)};
\fill[fci] (0,1.25) rectangle (8,2.45);
\fill[fci] (0,-1.15) rectangle (8,0.05);
\node at (4,2.05) {FCI, $\nu=2/3$ (spin and valley polarized)};
\node at (4,-0.75) {FCI, $\nu=2/3$};
\foreach \a/\b in {0/2,4/6}{\fill[sc] (\a,0.05) rectangle (\b,1.25);}
\foreach \a/\b/\l in {1/S,5/S}{\node at (\a,0.65) {$\l$: $P_{11}$, $A$};}
\foreach \a in {3,7}{\node at (\a,0.65) {$W$: $T_{11}$, $B$};}
\draw[chg] (0.1,1.12) -- (7.9,1.12); \draw[neu] (7.9,0.98) -- (0.1,0.98);
\draw[chg] (7.9,0.18) -- (0.1,0.18); \draw[neu] (0.1,0.32) -- (7.9,0.32);
\node[anchor=east] at (0,1.05) {$u$}; \node[anchor=east] at (0,0.25) {$\ell$};
\foreach \x/\j in {2/1,4/2,6/3}{
  \draw[densely dotted] (\x,0.05) -- (\x,1.25);
  \fill (\x,0.65) circle (1.6pt);
  \node[anchor=south] at (\x,1.25) {$\alpha_{\j}$};}
\begin{scope}[yshift=-4.1cm]
\node[anchor=west] at (-0.35,2.75) {(b)};
\fill[fci] (0,1.25) rectangle (8,2.45);
\fill[fci] (0,-1.15) rectangle (8,0.05);
\fill[sc] (0,0.05) rectangle (3.2,1.25); \fill[sc] (4.8,0.05) rectangle (8,1.25);
\node at (1.6,0.65) {$S_1$}; \node at (4,0.65) {$W$}; \node at (6.4,0.65) {$S_2$};
\draw[chg] (0.1,1.12) -- (7.9,1.12); \draw[neu] (7.9,0.98) -- (0.1,0.98);
\draw[chg] (7.9,0.18) -- (0.1,0.18); \draw[neu] (0.1,0.32) -- (7.9,0.32);
\foreach \x in {3.2,4.8}{\draw[densely dotted] (\x,0.05) -- (\x,1.25); \fill (\x,0.65) circle (1.6pt);}
\draw[thick,-{Latex[length=1.8mm]}] (3.2,1.3) .. controls (3.6,2.3) and (4.4,2.3) .. (4.8,1.3);
\draw[thick,-{Latex[length=1.8mm]}] (3.2,0.0) .. controls (3.6,-1.0) and (4.4,-1.0) .. (4.8,0.0);
\node[anchor=south] at (4,2.05) {$q_u$};
\node[anchor=north] at (4,-0.75) {$q_\ell$};
\node[anchor=west,align=left] at (0.12,1.85) {fragment:\\ $q_u=q_\ell=+e/3$};
\node[anchor=west,align=left] at (5.6,-0.5) {neutral:\\ $q_u=-e/3$,\\ $q_\ell=+e/3$};
\end{scope}
\end{tikzpicture}%
}
\caption{(a) Trench between two spin- and valley-polarized $\nu=2/3$ FCI regions, with edges $u$ and $\ell$. Each edge carries a downstream charge mode (solid) and an upstream neutral mode (dashed). Superconducting regions $S$ pin the pairing term $P_{11}$ and the neutral channel $A$; tunneling regions $W$ pin $T_{11}$ and the neutral channel $B$ (the region-by-region limit of Sec.~\ref{sec:zeromodes}). The walls bind $\mathbb{Z}_3$ zero modes $\alpha_j$. (b) A short junction $S_1|W|S_2$. Virtual tunneling of quasiparticles between the walls through the FCI bulk on each side of the trench, each returning to the edge it left, transfers, for example, a $2e/3$ fragment ($e/3$ on each edge) or no net charge (an $e/3$ quasihole on $u$ and an $e/3$ quasiparticle on $\ell$); $q_u$ and $q_\ell$ are the charges moved from $S_1$ to $S_2$. No such process changes the $\mathbb{Z}_3$ label, the eigenvalue of the fragment string between the interiors of $S_1$ and $S_2$ (Sec.~\ref{sec:josephson}).}
\label{fig:geometry}
\end{figure}

The construction requires two $\nu=2/3$ edges close enough for electrons to tunnel
between them and for Cooper pairs to form across them. Narrow superconducting
electrodes with quantum Hall edges on both sides have been
fabricated~\cite{Lee2017,Gul2022}, and the variant relevant here, in which the
superconducting region is defined by a gate inside the FQAH material, was studied
theoretically in Ref.~\cite{Bollmann2026}. We take the superconducting region to be
narrower than the induced coherence length, label the upper and lower edges $u$ and
$\ell$ [Fig.~\ref{fig:geometry}(a)], and reserve $t$ for the tunneling amplitude. Following
Ref.~\cite{Bollmann2026}, once the neutral sector is gapped the remaining charge
fields are
\begin{equation}
\Theta_\rho=\tfrac12(\phi_{u,\rho}-\phi_{\ell,\rho}),\qquad
\Phi_\rho=\tfrac12(\phi_{u,\rho}+\phi_{\ell,\rho}),
\end{equation}
with
\begin{equation}
[\partial_x\Theta_\rho(x),\,\Phi_\rho(y)]=i\frac{2\pi}{3}\,\delta(x-y).
\label{eq:commutator}
\end{equation}
The factor $2\pi/3$ reflects the $\nu=2/3$ topological order and is responsible for
the $\mathbb{Z}_3$, as opposed to $\mathbb{Z}_2$, character of the zero modes.

Electron tunneling and induced pairing bosonize to
\begin{equation}
H_{\rm pert}=\int\! dx\,\Bigl[-t(x)\cos(3\Theta_\rho)
+\Delta(x)\cos\bigl(3\Phi_\rho+\varphi\bigr)\Bigr],
\label{eq:perturbations}
\end{equation}
where $\varphi$ is the superconducting phase. Each term alone can gap the pair of
edge modes, as is most easily seen in the reduced basis of
Ref.~\cite{Bollmann2026},
$K_{\rm red}=\bigl(\begin{smallmatrix}0&3\\3&0\end{smallmatrix}\bigr)$, in which the
two terms correspond to $\Lambda_1=(1,0)$ and $\Lambda_2=(0,1)$. Both are Haldane
null vectors~\cite{Haldane1995},
\begin{equation}
\Lambda_1^{\mathsf T}K_{\rm red}\Lambda_1=\Lambda_2^{\mathsf T}K_{\rm red}\Lambda_2=0,
\end{equation}
but they cannot be pinned simultaneously, since
\begin{equation}
\Lambda_1^{\mathsf T}K_{\rm red}\Lambda_2=3\neq0 .
\label{eq:nonlocal}
\end{equation}
This incompatibility, with the value $3$ fixed by $\nu=2/3$, makes the two gapped
phases topologically distinct and binds $\mathbb{Z}_3$ zero modes to the walls
between them. Any candidate pairing term has to satisfy
Eq.~\eqref{eq:nonlocal} (Sec.~\ref{sec:pairing}).

\subsection{Scaling dimensions}
\label{sec:RG}

Treating the charge sector as a Luttinger liquid with parameter $K_\rho$, the
tree-level scaling dimensions of the two terms in Eq.~\eqref{eq:perturbations} are
\begin{equation}
\Delta_\Theta=\tfrac32 K_\rho,\qquad
\Delta_\Phi=\frac{3}{2K_\rho},
\label{eq:dims}
\end{equation}
for a static superconducting phase. Which of the two operators carries $K_\rho$ and
which $1/K_\rho$ is a matter of field normalization (rescaling to the standard form
$[\partial_x\theta,\phi]=i\pi\delta$ exchanges them), and none of the results below depends on this choice (we checked them in both conventions). If $\varphi$
fluctuates with stiffness $K_\varphi$, the second dimension becomes
$\Delta_\Phi=1/(2K_\varphi)+3/(2K_\rho)$~\cite{Bollmann2026}. A perturbation of scaling
dimension $\Delta$ in $1{+}1$ dimensions flows as
\begin{equation}
\frac{d g}{d\ell}=(2-\Delta)\,g ,
\label{eq:RG}
\end{equation}
and is relevant for $\Delta<2$.

The product $\Delta_\Theta\Delta_\Phi=9/4$ is independent of $K_\rho$. The two
operators are exchanged by $K_\rho\to1/K_\rho$, with the self-dual point at
$K_\rho=1$, where each dimension equals $3/2$, and the two are simultaneously relevant only within the window
\begin{equation}
\tfrac34<K_\rho<\tfrac43 .
\label{eq:window}
\end{equation}
Outside this window one of them is irrelevant, and for weak bare couplings the
system flows to a single gapped phase with no competing phase and hence no
domain-wall zero modes. (A sufficiently strong bare coupling can still pin the
field, since tree-level relevance is a weak-coupling statement.) Because $K_\rho$ is
set by the ratio of inter-edge to intra-edge Coulomb interaction, it depends on the
trench width and on screening, and Eq.~\eqref{eq:window} becomes a constraint on the
device design. Decoupled edges sit at $K_\rho=1$, and the allowed range of inter-edge coupling is derived in Sec.~\ref{sec:coulomb}.

\subsection{Domain-wall zero modes}

Let $\Delta(x)$ and $t(x)$ alternate along the trench. In pairing-dominated regions
$3\Phi_\rho+\varphi$ is pinned and $\Phi_\rho$ takes one of three values spaced by
$2\pi/3$; in tunneling-dominated regions $\Theta_\rho$ is pinned to one of three
values in the same way. By Eq.~\eqref{eq:commutator}, $e^{i\Phi_\rho}$ shifts
$\Theta_\rho$ by $2\pi/3$ and $e^{i\Theta_\rho}$ shifts $\Phi_\rho$ by the same
amount, so the zero mode at the $j$th wall involves both fields. Schematically,
\begin{equation}
\alpha_j\;\sim\;\exp\!\bigl[i\bigl(\Phi_\rho\pm\Theta_\rho\bigr)\bigr]\Big|_{x_j},
\label{eq:alpha}
\end{equation}
is the operator that advances the pinned value on one side of the wall by one step.
Equation~\eqref{eq:alpha} is not exact: $\Phi_\rho$ and $\Theta_\rho$ are pinned in
different regions, only one of them is pinned at $x_j$, and the commutator at
coinciding points is ambiguous. The proper construction builds $\alpha_j$ from the
pinned configurations on either side, with a factor $e^{-i\varphi_j/3}$ from the superconducting phase. We take the explicit form from
Refs.~\cite{Clarke2013,Bollmann2026} and use only the resulting algebra. Since the
fields are pinned in steps of $2\pi/3$, $\alpha_j^3=1$, and
Eq.~\eqref{eq:commutator} gives
\begin{equation}
\alpha_i\alpha_j=e^{i\frac{2\pi}{3}\,\mathrm{sgn}(j-i)}\,\alpha_j\alpha_i .
\label{eq:Z3}
\end{equation}
An operator built from a single field commutes with itself at all positions and
cannot satisfy Eq.~\eqref{eq:Z3}, which is why both $\Phi_\rho$ and $\Theta_\rho$
are needed.

For a Laughlin $\nu=1/m$ trench, keeping track of the fermion parity of the
condensate enlarges the algebra to $\mathbb{Z}_{2m}$~\cite{Clarke2013}. This does
not happen for $\nu=2/3$, where the operator that measures fermion parity is
condensed on both sides of the wall and the protected count is the same with or
without fermion parity (Sec.~\ref{sec:zeromodes}). Each wall has quantum dimension
$\sqrt3$, and $2N$ walls at fixed total charge have a ground-state degeneracy of
$3^{N-1}$~\cite{AliceaFendley2016}.

\subsection{Stability away from the self-dual point}
\label{sec:selfdual}

The $\mathbb{Z}_3$ parafermion critical theory, the three-state Potts point with
central charge $c=4/5$, is reached at the self-dual point $t=\Delta$, $K_\rho=1$~\cite{Mong2014,LGN2002}.
The zero modes, which sit at walls between two gapped phases, do not require this
tuning. In the related coupled-wire construction, Vaezi~\cite{Vaezi2014} emphasizes that the state is gapped, and hence
stable, as long as the deviation from self-duality, proportional to $|t-\Delta|$,
stays below the bulk gap. Consistently, the DMRG phase diagram of
Ref.~\cite{Bollmann2026} shows the relevant phases over extended regions of
parameter space.

\section{Ingredients that carry over}
\label{sec:transfers}

\subsection{Interface transparency}

Tunneling through an hBN spacer between the FCI and the superconductor would be
exponentially suppressed, but a spacer is not needed: one-dimensional edge
contacts~\cite{Wang2013} connect a metal directly to the exposed edge of an
encapsulated van der Waals stack. Lee \textit{et al.}~\cite{Lee2017} contacted
graphene quantum Hall edges with NbN and fitted their data with a
Blonder--Tinkham--Klapwijk barrier parameter $Z=0.18$, corresponding to a
transparency
\begin{equation}
\mathcal{T}=\frac{1}{1+Z^2}\simeq0.97 .
\end{equation}
Supercurrents have been carried through the quantum Hall regime~\cite{Amet2016},
Andreev reflection has been observed in the fractional quantum Hall
regime~\cite{Gul2022} and crossed Andreev reflection in a quantum anomalous Hall
insulator~\cite{Uday2024}, and superconducting junctions spanning about $20$
moir\'e cells have been made in twisted MoTe$_2$~\cite{Jia2025}.

A well-defined induced pairing gap on a fractional edge has, as far as we know, not
been demonstrated in any material, and no reliable value of $\Delta_{\rm ind}$ is
available for an FQAH edge. We regard this as the more serious materials problem
and treat $\Delta$ in Eq.~\eqref{eq:perturbations} as a phenomenological parameter
whose size we do not predict.

\subsection{Intrinsic superconductivity}

Superconductivity has been found in the first moir\'e Chern band of twisted bilayer
MoTe$_2$ at fillings adjacent to the FQAH states~\cite{Xu2025SC}. This makes
possible a device in which the superconducting and FQAH regions are defined by gates
on the same crystal, without a heterointerface and hence without the disorder,
lattice mismatch and work-function offsets that come with an external superconductor. The finger-gate geometry of Sec.~\ref{sec:geometry} is suited to such a device. In rhombohedral graphene, chiral superconductivity has been reported
inside a spin- and valley-polarized quarter-metal phase~\cite{Han2025}, and twisted
rhombohedral multilayers show superconductivity consistent with spin-triplet
pairing in the same devices as fractional high-Chern insulators~\cite{Huo2026}. Superconductivity obtained by doping an FQAH insulator has been studied theoretically~\cite{ShiSenthil2025}. Lotri\v{c} and Simon argue that the transition from the FCI to the superconductor in twisted MoTe$_2$ is probably not direct, with a reentrant integer quantum Hall phase in between~\cite{LotricSimon2026}. In a gate-defined device the density profile between the superconducting and FQAH regions must then avoid leaving an integer Chern strip along the trench, which would add edge modes to those considered here. The
pairing in these superconductors is probably not $s$-wave singlet
(Sec.~\ref{sec:pairing}).

\subsection{Energy scales}
\label{sec:energy}

The FQAH states themselves are well
established~\cite{Cai2023,Zeng2023,Park2023,Xu2023,Lu2024}. The value of $7$~meV
often quoted for the $\nu=2/3$ gap in twisted MoTe$_2$ is a local value of the
thermodynamic gap obtained from magnetic imaging~\cite{Redekop2024} in a sample
with considerable spatial disorder. Transport activation energies are smaller and
give a better scale for design estimates. In the first device with dissipationless
FCI transport (quantized Hall resistance and vanishing longitudinal resistance), the FCI gap of the $\nu=-2/3$ state is estimated at about $20$~K, or $1.7$~meV, with a spin gap of about $55$~K~\cite{ParkNP2026}.

These numbers give rough scales for the trench. Once the neutral sector is gapped, a pairing term of bare amplitude $\Delta_0$ at a cutoff $\Lambda$ set by the FCI gap grows according to Eq.~\eqref{eq:RG} until it is of order $\Lambda$, which gives a gap $E_g\sim\Lambda(\Delta_0/\Lambda)^{1/(2-\Delta_\Phi)}$, i.e., $E_g\sim\Delta_0^2/\Lambda$ for decoupled edges ($\Delta_\Phi=3/2$). For $\Lambda\approx1.7$~meV and assumed bare amplitudes $\Delta_0=0.1$ and $0.5$~meV this gives $E_g\approx6~\mu$eV and $0.15$~meV. With an assumed edge velocity $v\sim10^4$~m/s the zero modes extend over $\xi\sim\hbar v/E_g$, about $1~\mu$m and $45$~nm, respectively. These tree-level estimates assume a neutral gap of order $\Lambda$ and are upper bounds otherwise, but they indicate that the segments must be micrometers long unless the induced pairing is a sizable fraction of the FCI gap, and that dynamical probes must be slow on the scale $\hbar/E_g$.

\section{Pairing and tunneling on a spin-polarized trench}
\label{sec:fails}

Equation~\eqref{eq:perturbations} presupposes a gapped neutral sector. For the spin-polarized Jain state, Ref.~\cite{Bollmann2026} notes that the exact mechanism by which $\mathbb{Z}_3$ parafermions emerge ``remains elusive and is a subject of further research'', identifies the gapping of the neutral field as the most critical step, and assumes that ``some unspecified mechanism (disorder, interactions, or tunnelling/pairing among the trench)'' has gapped the neutral modes. Here we work out which of these processes select which neutral gap. We determine which operators exist
on a polarized trench, which neutral gaps they are compatible with, and which they
generate. The analysis uses only the edge $K$-matrix, the charge vector and the
requirement of zero conformal spin, so every statement reduces to integer linear
algebra and can be checked directly.

\subsection{Local operators}
\label{sec:neutral}

On a single edge we write vertex operators as $e^{i\bm{l}\cdot\bm{\phi}}$ with
$K=\mathrm{diag}(1,-3)$. The electron operators of the two modes, $\psi_1$ with
$\bm{l}=(1,0)$ and $\psi_2$ with $\bm{l}=(0,-3)$, both carry charge $1$ and have
conformal spins $\tfrac12$ and $-\tfrac32$. Their ratio $N=\psi_1^\dagger\psi_2$,
with $\bm{l}=(-1,-3)$, is a neutral boson, $\bm{t}^{\mathsf T}K^{-1}\bm{l}=0$, of
conformal spin
\begin{equation}
\tfrac12\,\bm{l}^{\mathsf T}K^{-1}\bm{l}=\tfrac12\bigl(1-3\bigr)=-1 .
\label{eq:spin}
\end{equation}
No nonzero integer $\bm{l}$ has zero conformal spin on a single edge, as this would
require $3l_1^2=l_2^2$, so a single edge admits no uniform gapping term. Random
tunneling between its two modes, which produces the KFP fixed point, equilibrates
the neutral mode without gapping it~\cite{KFP1994}, consistent with the upstream
neutral mode seen in experiments~\cite{Sabo2017,Bhattacharyya2019} and with the
two-mode structure obtained by combining counter-propagating $\nu=1$ and $\nu=1/3$
states~\cite{Cohen2019}. With disordered superconducting proximity, a single
$\nu=2/3$ edge has a family of disorder-stabilized fixed
points~\cite{TamChenLian2026}.

The two edges of the trench have opposite chirality, so all conformal spins change
sign on the lower edge. The local operators of the trench are generated by
$e_{1u}$, $e_{1\ell}$, $N_u$ and $N_\ell$, with $e_1$ the electron in the $\nu=1$
mode, and in this basis the form
$\langle\bm{l},\bm{l}'\rangle=\bm{l}^{\mathsf T}K^{-1}\bm{l}'$ reads
\begin{equation}
G=\begin{pmatrix}
1&0&-1&0\\
0&-1&0&1\\
-1&0&-2&0\\
0&1&0&2
\end{pmatrix},\qquad \det G=9 .
\label{eq:gram}
\end{equation}
An operator $v$ has conformal spin $\langle v,v\rangle/2$, and two operators at separated points commute up to the phase $e^{i\pi\langle v,v'\rangle}$. We call $v$ and $w$ compatible if $\langle v,w\rangle=0$, i.e., if their arguments commute, so that both cosines can be pinned at the same time; for $\langle v,w\rangle\neq0$ they cannot, even when the phase $e^{i\pi\langle v,w\rangle}$ equals one.

Table~\ref{tab:operators} lists the two-electron operators built from the mode
electrons $\psi_1$ and $\psi_2$ that add one electron to each edge or transfer one
electron across the trench. Four of them have zero conformal spin and can appear as
uniform terms, namely the pairing terms $P_{11}=\psi_{u1}\psi_{\ell1}$ and
$P_{22}=\psi_{u2}\psi_{\ell2}$ and the tunneling terms
$T_{11}=\psi_{u1}^\dagger\psi_{\ell1}$ and $T_{22}=\psi_{u2}^\dagger\psi_{\ell2}$.
The mixed pairing terms $\psi_{u1}\psi_{\ell2}$ and $\psi_{u2}\psi_{\ell1}$ have
conformal spin $\pm2$ and carry momentum. Operators dressed with additional neutral
excitations, such as $T_{11}-N_\ell$, can also have zero conformal spin, and they appear in Sec.~\ref{sec:zeromodes}.

\begin{table}[tbp]
\caption{Two-electron operators on a spin-polarized $\nu=2/3$ trench, in the
basis $(e_{1u},e_{1\ell},N_u,N_\ell)$. The last column gives the neutral channel that each operator is compatible with (Sec.~\ref{sec:channels}). The mixed tunneling operators $\psi_{u1}^\dagger\psi_{\ell2}$ and $\psi_{u2}^\dagger\psi_{\ell1}$, $(-1,1,0,1)$ and $(-1,1,-1,0)$, have spin $\pm2$ and are not listed.}
\label{tab:operators}
\begin{ruledtabular}
\begin{tabular}{lccc}
Operator & Vector & Spin & Neutral channel\\
\hline
$P_{11}=\psi_{u1}\psi_{\ell1}$ & $(1,1,0,0)$ & $0$ & $A$\\
$P_{22}=\psi_{u2}\psi_{\ell2}$ & $(1,1,1,1)$ & $0$ & $A$\\
$\psi_{u1}\psi_{\ell2}$ & $(1,1,0,1)$ & $2$ & none\\
$\psi_{u2}\psi_{\ell1}$ & $(1,1,1,0)$ & $-2$ & none\\
$T_{11}=\psi_{u1}^\dagger\psi_{\ell1}$ & $(-1,1,0,0)$ & $0$ & $B$\\
$T_{22}=\psi_{u2}^\dagger\psi_{\ell2}$ & $(-1,1,-1,1)$ & $0$ & $B$\\
$A=N_uN_\ell$ & $(0,0,1,1)$ & $0$ & $A$\\
$B=N_u^\dagger N_\ell$ & $(0,0,-1,1)$ & $0$ & $B$\\
\end{tabular}
\end{ruledtabular}
\end{table}

\subsection{Neutral channels and compatibility}
\label{sec:channels}

Table~\ref{tab:operators} gives
\begin{equation}
P_{22}-P_{11}=A,\qquad T_{22}-T_{11}=B ,
\label{eq:generated}
\end{equation}
with $A=N_uN_\ell$ and $B=N_u^\dagger N_\ell$. Both have zero charge and zero
conformal spin, and each can gap the two neutral modes together. These are the two
channels identified by Park \textit{et al.}~\cite{Park2024}, neutral-mode
superconductivity and neutral-mode backscattering; in the coordinates of their
Eqs.~(11)--(13), $A=-\bm{M}_{\rm sup}$ and $B=\bm{M}_{\rm back}$. They are incompatible, $\langle A,B\rangle=4$, and each charge term is compatible with only one of them,
\begin{equation}
\begin{aligned}
&\langle P_{11},A\rangle=0,&&\langle P_{11},B\rangle=2,\\
&\langle T_{11},B\rangle=0,&&\langle T_{11},A\rangle=2 .
\end{aligned}
\label{eq:compat}
\end{equation}
If the neutral modes are gapped in channel $A$, the argument of the pairing term commutes with the pinned field and the term reduces to $\Delta\cos(3\Phi_\rho+\varphi)$ up to a constant factor, whereas the argument of the bare tunneling term does not, and the term averages to zero. In channel $B$ the roles are reversed. Single electrons are gapped in both
channels, since every operator of odd charge has odd $\langle\cdot,A\rangle$ and
$\langle\cdot,B\rangle$, in agreement with the effective theory of
Ref.~\cite{Park2024}, in which the smallest local charge is $2$.

In the unpolarized $(1,1,2)$ state the situation is different. The singlet pairing
terms are then $\psi_{u\uparrow}\psi_{\ell\downarrow}$ and
$\psi_{u\downarrow}\psi_{\ell\uparrow}$, the tunneling terms are
$\psi_{u\uparrow}^\dagger\psi_{\ell\uparrow}$ and
$\psi_{u\downarrow}^\dagger\psi_{\ell\downarrow}$, and the neutral operator
generated by the difference of the two pairing terms coincides, up to sign, with
the one generated by the difference of the two tunneling terms. Singlet pairing and
tunneling thus select the same neutral channel, in line with the statement of
Ref.~\cite{AliceaFendley2016} that both gap the spin sector in the same way and
with the remark of Ref.~\cite{Park2024} that for spin-unpolarized edges with
conserved spin only the backscattering channel is available. The difference between the two cases comes from the spin structure: singlet pairing joins opposite spins, which reverses the sign of the neutral content on one edge, and equal-spin pairing does not. Since the two states have the same topological order, the operators of the $(1,1,2)$ construction also exist on the polarized trench. In the basis $(e_1,\,e_1-N)$ the Gram matrix of one polarized edge is exactly $K'$, so the spin-up electron corresponds to $\psi_1$ and the spin-down electron to $\psi_1^2\psi_2^\dagger$, a charge-one operator of conformal spin $\tfrac12$ built from three electron operators. The singlet pair $\psi_{u\uparrow}\psi_{\ell\downarrow}$ then becomes the dressed operator $P'=P_{11}-N_\ell$ introduced in Sec.~\ref{sec:zeromodes}, and singlet pairing and tunneling select channel $B$. Spin polarization thus changes which operators are low order, not the topological order: on a polarized edge the bare mode electrons are the low-order operators, while the analog of the singlet pair is a higher-order composite. Conversely, equal-spin pairing in the $(1,1,2)$ state selects a different channel from tunneling, as in the polarized case.

\subsection{Neutral gaps generated by pairing and tunneling}
\label{sec:generated}

By Eq.~\eqref{eq:generated}, the charge-sector processes generate a neutral cosine
at second order. The two pairing terms give $g_A\propto\Delta_{22}\Delta_{11}^*$
times $A+\mathrm{h.c.}$ in a pairing region, and the two tunneling terms give
$g_B\propto t_{22}t_{11}^*$ times $B+\mathrm{h.c.}$ in a tunneling region, where
$\Delta_{aa}$ and $t_{aa}$ are the amplitudes of $P_{aa}$ and $T_{aa}$. Both amplitudes of a kind must be nonzero. Pairing that acts only on the $\nu=1$ modes, for example, generates no neutral term.

The inter-edge Coulomb interaction is a further source. It contains products of
the intra-edge operators $N_u$ and $N_\ell$ and produces both $A$ and $B$ directly,
at first order~\cite{Park2024}. The channel that wins is decided by the
competition between these Coulomb terms and the terms generated by pairing and
tunneling, and by their scaling dimensions.

When charge and neutral modes are decoupled on each edge, as at the KFP fixed point, and the inter-edge interaction couples only the charge densities, $A$ and $B$ keep scaling dimension $2$ for any strength of this coupling (Table~\ref{tab:dims}), i.e., they are marginal. A density--density coupling between the neutral modes of the two edges splits the two dimensions, making one channel relevant and the other irrelevant. The neutral-sector problem at this point was analyzed by
V\"ayrynen, Goldstein and Gefen~\cite{Vayrynen2022} for two counter-propagating
copies of the $\nu=2/3$ neutral mode. The neutral sector of each edge has an
SU(2)$_1$ symmetry~\cite{KFP1994}, $A$ and $B$ are the products $J^+J^+$ and
$J^+J^-$ of the SU(2) currents of the two edges, each of dimension $2$, and the
one-loop flow of the current--current couplings~\cite{Giamarchi2003} is of
Kosterlitz--Thouless type, with the density--density coupling of the neutral modes selecting the channel that becomes relevant. This coupling arises from the Coulomb interaction because neutral excitations are charge dipoles; Ref.~\cite{Vayrynen2022} argues that it is weakly attractive and favors pairing, with a gap that scales as a large power of the bare coupling. Away from the KFP point the dimensions depend on the full interaction
matrix~\cite{Park2024}. In the cases we examined, a charge--neutral coupling within each edge raises both dimensions above $2$, and an additional repulsive inter-edge coupling of the charge densities then favors backscattering over superconductivity. With both dimensions above $2$, a neutral gap at weak coupling requires the KFP point, strong bare coupling or disorder. The KFP fixed point is
reached through random tunneling, which rotates the SU(2) frame of each edge
randomly along the trench. The random part of the coupling then reduces to the
localization problem of Ref.~\cite{Park2024}, and the uniform argument applies only
to what survives the averaging. The same rotation acts on the charge terms. In the basis $(e_1,\,e_1-N)$ of Sec.~\ref{sec:channels}, the charge-one operators $e_1$ and $e_1-N$ of an edge form an SU(2) doublet at the KFP point, and $P_{11}$ and the dressed pair $P'$ belong to one multiplet of the two edges, so random rotations mix pairing terms that select $A$ and $B$. The selection rules of Eq.~\eqref{eq:compat} thus apply to the bare operators and on scales shorter than the length over which random tunneling rotates the frame, where the edge has not yet reached the KFP point. On longer scales the channel is set by disorder, which leads to one of the limits of Sec.~\ref{sec:zeromodes}, and the protected count is unchanged. For disorder correlated between the two edges,
$\xi_u=\xi_\ell^*$ in the amplitudes of the random intra-edge tunneling, the SU(2)
rotation that removes the disorder leaves the neutral pairing coupling invariant;
for uncorrelated disorder it does not~\cite{Vayrynen2022}.

\emph{Momentum.} Zero conformal spin is necessary for a uniform term but not
sufficient, since the Fermi-momentum factors of the electron operators must also
cancel. Let the modes of the upper edge cross the Fermi level at $k_1$ and $k_2$,
and suppose the lower edge is related to it by a symmetry that exchanges the edges
and reverses their chirality, so that its modes cross at $-k_1$ and $-k_2$. In a
valley-polarized system this symmetry can be neither a twofold rotation about the
out-of-plane axis, which exchanges the valleys, nor a mirror alone, which reverses
the Hall conductance. The product of time reversal and an in-plane reflection that
maps the valley momentum $K$ to $-K$ preserves the valley, the spin polarization
and the Hall conductance, and plays this role for a trench along the reflection
line whenever the moir\'e crystal has such a symmetry. Under this assumption
$P_{11}$, $P_{22}$ and $A$ carry zero momentum, $T_{11}$ and $T_{22}$ carry $2k_1$
and $2k_2$, and $B$ carries $2(k_1-k_2)$. In a clean trench the pairing terms and
the neutral channel they select conserve momentum, while tunneling and the
backscattering channel require a momentum source such as disorder or a
commensurate moir\'e wave vector. Reference~\cite{Vayrynen2022} made this observation for the neutral operators, and it extends here to the charge terms that select each channel. It favors channel $A$, the first of the
two limits discussed in Sec.~\ref{sec:zeromodes}, on the scales where the selection rules apply.

\subsection{Zero-mode structure}
\label{sec:zeromodes}

Gapped edges correspond to maximal sets of mutually compatible local operators of zero conformal spin~\cite{Haldane1995,Levin2013}, and the protected degeneracy at
the walls between two gapped regions is determined by the anyons that each region
condenses~\cite{BarkeshliJianQi2013,Kapustin2014}. Let $\Gamma=\mathbb{Z}^4$ be the
lattice of local operators with the form $G$ of Eq.~\eqref{eq:gram}, and
$\Gamma_b\subset\Gamma$ its sublattice of even fermion number, which contains all
physical perturbations. A region that pins the maximal isotropic sublattice $L$
condenses the subgroup
\begin{equation}
H(L)=\bigl(\mathrm{span}_{\mathbb{Q}}L\cap\Gamma_b^*+\Gamma_b\bigr)/\Gamma_b
\end{equation}
of the finite group $\Gamma_b^*/\Gamma_b$, with $\Gamma_b^*$ the dual
lattice with respect to $G$. For $2N$ walls between regions that alternately pin $L_1$ and
$L_2$, at fixed total charge and fermion parity, the protected degeneracy is $(d^2)^{N-1}$ with
\begin{equation}
d^2=\frac{|H(L_1)|}{|H(L_1)\cap H(L_2)|},
\label{eq:kapustin}
\end{equation}
$d$ being the quantum dimension of a wall~\cite{Kapustin2014,BarkeshliJianQi2013b}; we call $d^2$ the number of protected states per pair of walls.
The count of Ref.~\cite{Kapustin2014} uses the anyon data only and does not keep track of fermion parity. For fermionic states, Barkeshli, Jian and Qi note that the anyon data fix the defect degeneracy only up to a factor $\sqrt2$ from Majorana modes of the purely one-dimensional fermionic physics~\cite{BarkeshliJianQi2013,BarkeshliJianQi2013b}. We include fermion parity by working with $\Gamma_b$ instead of $\Gamma$. By spin--statistics the fermion parity of a local operator $\gamma$ is $\langle\gamma,\gamma\rangle$ mod 2, so $\Gamma_b$ is an even lattice, and $\Gamma_b^*/\Gamma_b$, of order $4|\Gamma^*/\Gamma|$, is the anyon group of the bosonic theory obtained by gauging fermion parity, to which the result of Ref.~\cite{Kapustin2014} applies. For $K=\mathrm{diag}(m,-m)$ the anyon data alone give $d=\sqrt m$, while $\Gamma_b$ gives $d=\sqrt{2m}$, the $\mathbb{Z}_{2m}$ parafermions of Refs.~\cite{Lindner2012,Clarke2013,Cheng2012}, so the construction supplies the factor $\sqrt2$. Equivalently, $d^2$ is the dimension of the irreducible representation of the string operators that cross the walls, restricted to strings in $\Gamma_b^*$, i.e., strings that braid trivially with every bosonic local operator; the two computations agree. A related treatment of a Laughlin--superconductor heterostructure describes the superconductor as a topological order with deconfined half-flux vortices, which also brings fermion parity into the anyon data~\cite{Yoshitome2025}. As a check, Eq.~\eqref{eq:kapustin} gives 2 for
superconductor/ferromagnet walls on a helical edge (Majorana modes), 6 for a
$\nu=1/3$ Laughlin trench (the $\mathbb{Z}_6$ parafermions of
Ref.~\cite{Clarke2013}; 3 if fermion parity is ignored), and 3 for the unpolarized
$(1,1,2)$ trench (the $\mathbb{Z}_3$ parafermions of
Refs.~\cite{Mong2014,AliceaFendley2016}).

For the polarized trench $|\Gamma_b^*/\Gamma_b|=36$, and each of the pinned sets
considered below is maximal, with $H\cong\mathbb{Z}_3\times\mathbb{Z}_2$. The
$\mathbb{Z}_3$ factor is generated by $(P_{11}+A)/3$, which adds charge $e/3$ to each edge, $2e/3$ in total, the fragment that carries the fractional
Josephson current (Sec.~\ref{sec:signatures}). The $\mathbb{Z}_2$ factor is the
neutral vector $A/2$. It creates a neutralon, a chargeless semion of conformal spin
$\mp\tfrac14$~\cite{Vayrynen2022}, on each edge, anticommutes with single
electrons, and hence measures fermion parity. We consider two limits.

\emph{Uniform neutral channel.} Suppose that a single channel, say $A$, is pinned
along the whole trench, for instance by disorder~\cite{Park2024} or because only
the pairing side conserves momentum (Sec.~\ref{sec:generated}). The tunneling
region must then use the dressed operator $T'=T_{11}-N_\ell$, which has zero conformal spin and is compatible with $A$. With $\{P_{11},A\}$ and $\{T',A\}$ pinned, the
two regions share $A/2$ and differ in the $\mathbb{Z}_3$ factor, and
Eq.~\eqref{eq:kapustin} gives 3 states per pair of walls, the $\mathbb{Z}_3$ zero
modes assumed in Ref.~\cite{Bollmann2026}. The dressed operator has the same
scaling dimension as $T_{11}$ after the neutral modes are gapped (before, $T'$ and $T_{11}$ have dimensions $\Delta_T+\Delta_A/4$ and $\Delta_T+\Delta_B/4$, equal when the neutral modes of the two edges are uncoupled),
but it involves an extra neutral excitation on one edge, and we expect its bare amplitude to be smaller. If $B$ is pinned everywhere instead, the pairing term has to be dressed, $P'=P_{11}-N_\ell$, and the count is again 3. By Sec.~\ref{sec:channels} this limit is the $(1,1,2)$ construction written in polarized variables.

\emph{Neutral gaps set region by region.} Suppose now that each region gaps its
neutral sector in the channel generated by its own processes
(Sec.~\ref{sec:generated}), so that $\{P_{11},A\}$ and $\{T_{11},B\}$ are pinned.
Although $A$ and $B$ are incompatible,
\begin{equation}
\tfrac12A-\tfrac12B=N_u\in\Gamma_b ,
\label{eq:sameanyon}
\end{equation}
so the two regions condense the same $\mathbb{Z}_2$ element and
Eq.~\eqref{eq:kapustin} again gives 3. In the neutral sector alone, whose anyons are
those of U(1)$_2\times$U(1)$_{-2}$, there is a single Lagrangian subgroup, so
neutral-mode superconductivity and neutral-mode backscattering are the same gapped
boundary up to local operators, and a change of channel at a wall binds no
protected zero mode. Counting all string operators that commute with the pinned terms, without the restriction to $\Gamma_b^*$, i.e., evaluating $|\det M|$ with $M_{ij}=\langle v_i,w_j\rangle$ for the
bases $\{v_i\}=\{P_{11},A\}$ and $\{w_j\}=\{T_{11},B\}$, gives 12 for this
configuration, but the additional states are not protected. Strings such as
$P_{11}/2$ fail to commute with bosonic local operators such as $N_u$ acting at the
walls, which can split them, and $A/2$ is condensed in both regions by
Eq.~\eqref{eq:sameanyon}.

Either limit thus gives $\mathbb{Z}_3$ parafermions. The $\mathbb{Z}_2$ factor that measures fermion parity is shared by the two regions, so that fixing
the fermion parity does not enlarge the count. In the Laughlin case the
corresponding factor is half of the pairing vector, which carries charge and is not
condensed in the tunneling region. The two limits differ only in energetic
respects: which charge term enters in bare and which in dressed form, and whether
the walls, which carry no protected mode associated with the neutral channel, bind
unprotected low-energy states.

\subsection{Coulomb interaction between the edges}
\label{sec:coulomb}

The Coulomb interaction between the edges couples the two charge modes. For intra-edge coupling $V$ and inter-edge coupling $U$ in the charge sector, and $\lambda=U/V$, the Luttinger parameter is $K_\rho=\sqrt{(1-\lambda)/(1+\lambda)}$,
and once the neutral modes are gapped the pairing and tunneling terms have scaling
dimensions
\begin{equation}
\Delta_P=\tfrac32\sqrt{\frac{1+\lambda}{1-\lambda}},\qquad
\Delta_T=\tfrac32\sqrt{\frac{1-\lambda}{1+\lambda}},
\label{eq:coulomb}
\end{equation}
whose product is $9/4$, as in Sec.~\ref{sec:RG}. Decoupled edges, $\lambda=0$, give
$3/2$ for both, i.e., the trench sits at the self-dual point. Both terms are
relevant for
\begin{equation}
|\lambda|<\tfrac{7}{25},
\label{eq:lambda}
\end{equation}
which is the window~\eqref{eq:window} in the normalization of
Ref.~\cite{Bollmann2026}; $\lambda=7/25$ and $\lambda=-7/25$ correspond to
$K_\rho=3/4$ and $K_\rho=4/3$. A repulsive inter-edge coupling ($\lambda>0$) makes
tunneling more relevant and pairing less relevant. The trench should thus be narrow
enough for sizable $\Delta$ and $t$ but screened, e.g., by a nearby gate, so that
the inter-edge Coulomb coupling stays below $0.28$ of the intra-edge coupling once the neutral gap has formed. We
verified Eq.~\eqref{eq:coulomb} by diagonalizing the full four-mode theory with
charge and neutral velocities and an inter-edge coupling of the charge densities
(Table~\ref{tab:dims}).

These dimensions assume that the neutral sector is already gapped. Before it is, $P_{11}$ and $T_{11}$ also contain the neutral parts $A/2$ and $B/2$ of their electron operators, which add $\Delta_A/4$ and $\Delta_B/4$ to their dimensions. When the neutral modes of the two edges are uncoupled this is $\tfrac12$ each (Table~\ref{tab:dims}): both terms are then marginal for decoupled edges, and a repulsive inter-edge coupling makes pairing irrelevant and tunneling relevant, with dimensions $2.16$ and $1.86$ at $\lambda=0.1$. A neutral coupling that favors channel $A$ lowers the pairing dimension; for $\Delta_A=1.81$, obtained with an attractive inter-edge density coupling of the neutral modes of $10\%$ of the neutral velocity, it is $1.95$ at $\lambda=0$, and pairing stays relevant up to $\lambda\approx0.03$. An attractive inter-edge charge coupling, $\lambda<0$, would also make pairing relevant, but the Coulomb interaction gives $\lambda>0$. Tunneling, on the other hand, carries momentum $2k_1$ in a clean symmetric trench (Sec.~\ref{sec:generated}), and at the KFP point both amplitudes vary randomly along the trench. At weak coupling neither charge term thus gaps a region reliably on its own, and the neutral gap has to form first, from the neutral coupling, from disorder, or from strong bare amplitudes. Equation~\eqref{eq:lambda} applies once it has formed; before that, pairing favors $\lambda\le0$.

\begin{table}[tbp]
\caption{Scaling dimensions from the four-mode theory as a function of
$\lambda=U/V$. $\Delta_P$ and $\Delta_T$ are the charge parts of $P_{11}$ and
$T_{11}$, which set their dimensions after the neutral modes are gapped; $\Delta^{(0)}_P$ and $\Delta^{(0)}_T$ are the full dimensions before the neutral modes are gapped. $A$ and $B$ are the neutral
operators. Charge and neutral modes are decoupled on each edge, and the neutral modes of the two edges are uncoupled.}
\label{tab:dims}
\begin{ruledtabular}
\begin{tabular}{ccccccc}
$\lambda$ & $\Delta_P$ & $\Delta_T$ & $\Delta_P\Delta_T$ & $\Delta^{(0)}_P$ & $\Delta^{(0)}_T$ & $\Delta_A=\Delta_B$\\
\hline
$0$ & $1.500$ & $1.500$ & $2.25$ & $2.000$ & $2.000$ & $2$\\
$0.10$ & $1.658$ & $1.357$ & $2.25$ & $2.158$ & $1.857$ & $2$\\
$0.20$ & $1.837$ & $1.225$ & $2.25$ & $2.337$ & $1.725$ & $2$\\
$0.28$ & $2.000$ & $1.125$ & $2.25$ & $2.500$ & $1.625$ & $2$\\
\end{tabular}
\end{ruledtabular}
\end{table}

\subsection{Which superconductors can supply the pairing}
\label{sec:pairing}

The FQAH state is spontaneously spin- and valley-polarized, which is why its Hall
effect is anomalous and appears at zero field. An $s$-wave singlet superconductor
pairs electrons of opposite spin, and a fully polarized edge has no opposite-spin
partners. Proximity coupling nevertheless remains possible. Nakai, Nomura and
Tanaka~\cite{Nakai2021} found that the pairing induced on a spin-polarized quantum
anomalous Hall edge by singlet leads is equal-spin triplet, with even- and odd-frequency components, and crossed Andreev reflection has been observed across a
narrow Nb electrode on such an edge~\cite{Uday2024}. Equal-spin Andreev reflection
between spin-resolved quantum Hall states and singlet superconductors has also been
reported~\cite{Matsuo2018}, and spin-triplet supercurrents carried by
spin-polarized Landau levels were predicted earlier~\cite{vanOstaay2011}.

On a polarized trench the pair operators $P_{11}$ and $P_{22}$ are equal-spin by
construction. For two electrons of the same spin and valley, fermionic
antisymmetry makes the static pair amplitude odd under exchange of their
coordinates, so it has odd orbital parity. The amplitudes are
\begin{equation}
\Delta_{aa}\propto\int\!dy\,dy'\,F(y,y')\,\chi^u_a(y)\,\chi^\ell_a(y'),
\label{eq:amp}
\end{equation}
where $\chi_a$ are the transverse profiles of the edge modes and $F$ is the static
anomalous function of the superconductor for two electrons with the spin and
valley of the FQAH state. This leads to four conditions.

(i) $F$ must contain equal-spin pairs with $S_z=\pm1$ along the polarization axis.
An $s$-wave singlet superconductor with only Ising spin--orbit coupling has none:
the singlet has $S_z=0$, and the triplet component generated by Ising coupling has
its $\bm{d}$-vector along $z$, which again means $S_z=0$, with the pair spin in the
plane~\cite{Zhou2016}. In this case $\Delta_{aa}$ vanishes at leading order, and a
spin flip would require intervalley scattering, since Ising coupling locks spin to
valley. A Rashba component or an inhomogeneous magnetization breaks $S_z$ conservation and allows a nonzero $\Delta_{aa}$; spin-flip processes of this kind underlie the equal-spin Andreev reflection observed in Ref.~\cite{Matsuo2018}. Proposals for spin-polarized Laughlin trenches rely on
this route, using the spin--orbit coupling of a superconductor such as
NbN~\cite{Nielsen2022}.

(ii) $F$ must be even in frequency. The static term in the edge theory is set by the anomalous self-energy at $\omega\to0$, and an odd-frequency amplitude, $F(\omega)=-F(-\omega)$, vanishes there; its equal-time value $\int\!d\omega\,F(\omega)$
vanishes as well. The odd-frequency part of the induced pairing does not contribute to a
static term, although it is present at unequal times and can affect dynamical
measurements.

(iii) The pairing must be intravalley, in the valley of the FQAH state. An intravalley pair carries the crystal momentum $2K\equiv-K$, which is not a reciprocal lattice vector, so the amplitude supplied by a conventional superconductor with zero-momentum pairs is suppressed unless the contact transfers momentum on the atomic scale; this favors a superconductor in the same crystal. The chiral
intravalley $p+ip$ and $f+if$ states proposed~\cite{WangZaletel2025,XuZhang2025} for the superconductor observed next
to the FQAH state~\cite{Xu2025SC} satisfy (i) to (iii),
provided the superconducting and FQAH regions are polarized in the same valley.
The valley polarization of the FQAH state follows the sign of the training field; whether the superconducting region follows it in the same way remains to be established,
and a mismatch would suppress $\Delta_{aa}$. A Kohn--Luttinger analysis of flat
Chern bands likewise finds chiral $p$- and $f$-wave order for spin-polarized
pairing~\cite{Zou2026}.

(iv) The chirality of the $p+ip$ or $f+if$ state enters only through the phase of
$\Delta_{aa}$, which is absorbed into $\varphi$. In a trench whose edges are related
by the symmetry of Sec.~\ref{sec:generated}, $P_{11}$ and $P_{22}$ carry zero
momentum, because the two electrons occupy the same mode type on edges of opposite
chirality. An asymmetry $\delta k$ between the Fermi momenta of the two edges gives
the term momentum $\delta k$ and cuts off its growth beyond lengths of order
$1/\delta k$.

\subsection{Open questions}

Several issues are left open. An estimate of $\Delta_{11}$ and $\Delta_{22}$ for a given superconductor requires the
edge-mode profiles and the anomalous function in Eq.~\eqref{eq:amp}. The
competition between the neutral channels generated in each region and
disorder-induced localization~\cite{Park2024,Yutushui2024} decides between the two
limits of Sec.~\ref{sec:zeromodes}, and hence which charge term must be dressed.
Unprotected states bound to walls at which the neutral channel changes do not alter
the protected degeneracy but can mask the zero-bias signatures of
Sec.~\ref{sec:signatures}. Before the neutral gap forms, pairing is marginal or irrelevant for repulsive inter-edge coupling unless an attractive neutral coupling favors its channel (Sec.~\ref{sec:coulomb}), so whether a pairing region gaps its neutral sector at realistic couplings is the main open problem; it requires treating the charge and neutral couplings together, beyond tree level. The time over which the $6\pi$ period of Sec.~\ref{sec:josephson} survives is set by quasiparticles entering the junction from the bulk, by processes reaching the far ends of the superconducting regions and by inelastic processes, and requires a microscopic estimate.

\section{Experimental signatures}
\label{sec:signatures}

Table~\ref{tab:signatures} collects the expected values discussed in this section.

\begin{table}[tbp]
\caption{Expected signatures for $\nu=2/3$ at $f=10$~GHz (at $1$~GHz, $V_1=6.2~\mu$V; the drive must satisfy $hf\ll E_g$, Sec.~\ref{sec:shapiro}). The Josephson and Shapiro values hold with or without fixed fermion parity and are not changed by processes confined to the junction when the tunneling segments are not FCI; inelastic processes, quasiparticles entering from the bulk and processes reaching the far ends of the superconducting regions restore $2\pi$ periodicity (Sec.~\ref{sec:josephson}). The thermal conductance refers to a trench tuned to the self-dual point.}
\label{tab:signatures}
\begin{ruledtabular}
\begin{tabular}{lc}
Quantity & Expected value \\
\hline
Josephson period & $6\pi$\\
Oscillations per $\Phi_0$ & $1/3$ \\
Shapiro $V_1$ & $62.0$~$\mu$V \\
Zero-bias $G$ & $4e^2/3h$ \\
Peak splitting & $3$ resonances \\
$\kappa$ along critical trench ($100$~mK) & ${\sim}7.6\times10^{-14}$~W/K \\
\end{tabular}
\end{ruledtabular}
\end{table}

\subsection{Fractional Josephson effect}
\label{sec:josephson}

The superconducting phase enters Eq.~\eqref{eq:perturbations} only through
$\cos(3\Phi_\rho+\varphi)$. Although the Hamiltonian is $2\pi$-periodic in
$\varphi$, the adiabatic ground state is not. For $\Phi_\rho$ compact with period
$2\pi$, the minima lie at
\begin{equation}
\Phi_\rho=\frac{2\pi n-\varphi}{3},\qquad n=0,1,2 ,
\end{equation}
three values spaced by $2\pi/3$. Advancing $\varphi$ by $2\pi$ moves each minimum
onto the next and leaves the system in a different member of the degenerate set,
so three such steps are needed to return to the original minimum,
\begin{equation}
\varphi\to\varphi+6\pi .
\end{equation}
Keeping the lowest harmonic, $E(\varphi)=-E_0\cos(\varphi/3)$ and
$I=(2e/\hbar)\partial_\varphi E\propto\sin(\varphi/3)$. The argument assumes that
$\Phi_\rho$ is compact and that the evolution is adiabatic; the second assumption fails in the presence of the inelastic processes discussed below. The $6\pi$ period agrees with Alicea and Fendley~\cite{AliceaFendley2016}. Exact diagonalization of a time-reversal-symmetric bilayer of $\pm1/3$ Laughlin states, a spin-unpolarized system in a geometry different from the trench junction considered here, finds a $6\pi$ period~\cite{Repellin2018}.

More generally, a degenerate set that is cyclically permuted by
$\varphi\to\varphi+2\pi$ and carries a $\mathbb{Z}_N$ label has period $2\pi N$.
For $N=2$ this is the $4\pi$ Majorana result, and the $\mathbb{Z}_3$ structure of
Eq.~\eqref{eq:perturbations} gives $6\pi$, with the current carried by
charge-$2e/3$ fragments~\cite{AliceaFendley2016}. For $\nu=2/3$ the protected label is $\mathbb{Z}_3$ whether or not the fermion parity is fixed (Sec.~\ref{sec:zeromodes}), so stray electrons, which change only the fermion parity, do not restore $2\pi$ periodicity~\cite{AliceaFendley2016}.

Processes confined to the junction cannot change the label either. The $\mathbb{Z}_3$ label is the eigenvalue of the string operator of the fragment $(P_{11}+A)/3$ that runs from the interior of $S_1$ to the interior of $S_2$ [Fig.~\ref{fig:geometry}(b)]. Its endpoints lie where the fragment is condensed, so the string creates no excitation, and its three eigenvalues distinguish the three minima of $\Phi_\rho$ in $S_2$ relative to those in $S_1$, which $\varphi\to\varphi+2\pi$ maps onto one another. Any process confined to the junction is a product of vertex operators $e^{i\bm{u}_k\cdot\bm{\phi}(x_k)}$ located between the endpoints of the string, with $\sum_k\bm{u}_k$ a local operator. Examples are the tunneling of the fragment across $W$, the neutral process in which an $e/3$ quasihole crosses $W$ on one edge and an $e/3$ quasiparticle on the other, and the exchange of single electrons with the superconductor. The sum is local because the edges run continuously through $W$, an insulating segment of the trench and not an FCI, so no quasiparticle can cross from $u$ to $\ell$ within the junction. With $[\bm{a}\cdot\bm{\phi}(x),\bm{b}\cdot\bm{\phi}(y)]=i\pi\langle\bm{a},\bm{b}\rangle\,\mathrm{sgn}(x-y)$, its commutation phase with the string is $\exp[2\pi i\langle(P_{11}+A)/3,\sum_k\bm{u}_k\rangle]=1$, because the fragment is a quasiparticle of the FQAH state and has an integer product with every local operator, single electrons included. The same holds in all three configurations of Sec.~\ref{sec:zeromodes}, with the fragment $(P'-B)/3$ when $P'$ is pinned (ancillary file \texttt{junction.py}). The label, and with it the $6\pi$ period, can thus be changed only by processes that reach the far ends of the superconducting regions, whose amplitude decays with their length (including the transfer of quasiparticles between $u$ and $\ell$ around the ends of the trench when both FCI regions belong to one sample), by quasiparticles entering the junction from the bulk, and by the inelastic processes discussed below. The inelastic processes are suppressed by the gaps of the superconducting regions, including the neutral gap, which may be the smaller one.

The same criterion applies to the other cases of Sec.~\ref{sec:zeromodes}. In a Laughlin $\nu=1/m$ trench the $\mathbb{Z}_m$ part of the $\mathbb{Z}_{2m}$ label is carried by the quasiparticle $P/m$ and is protected in the same way, whereas the remaining $\mathbb{Z}_2$, generated by $P/2$, changes sign under single electrons and is protected only when the fermion parity is fixed. We thus find $4m\pi$ at fixed fermion parity and $2m\pi$ otherwise, which for Majorana modes reduces to $4\pi$ and $2\pi$. For $\nu=2/3$ the whole label is carried by a quasiparticle, which is why the period is $6\pi$ with or without fixed fermion parity. Cao \textit{et al.}~\cite{Cao2024} find that a backscattering term at a junction between two Laughlin fractional topological superconductors, each described by its own fields, reduces the period to $4\pi$ at $T=0$, and to $2\pi$ at finite temperature unless the fermion parity is fixed. Our criterion reproduces this result if quasiparticles can pass around the ends of the superconducting regions within the junction, as when the FQH liquid fills the gap between two superconducting fingers and the edge of each finger wraps around its tip. Such a process crosses the fragment string, and the criterion then gives $4\pi$ at fixed fermion parity and $2\pi$ otherwise for $\nu=1/m$, and $2\pi$ for $\nu=2/3$. We attribute the difference between their $4\pi$ and the $4m\pi$ found here to this difference in edge geometry. The $6\pi$ period thus requires the tunneling segments to be part of the trench: in a gate-defined device, the gates must remove the FQAH state from $W$ as well as from $S$. How the Josephson periodicity depends on the type of defect in a $\nu=1/3$ Laughlin--superconductor heterostructure is discussed in Ref.~\cite{Yoshitome2025}. The $4m\pi$ of Refs.~\cite{Lindner2012,Clarke2013}, $12\pi$ for $m=3$, applies to Laughlin $\nu=1/m$ systems at fixed fermion parity, and it does not carry over to $\nu=2/3$.

In a SQUID with $\varphi=2\pi\Phi/\Phi_0$, a $6\pi$ period corresponds to one
oscillation per $3\Phi_0$, fewer than in a conventional junction. Higher harmonics
such as $\sin(3\varphi)$, which arise in transparent conventional junctions from
multiple Cooper-pair tunneling, leave the $2\pi$ period unchanged and carry no
topological information.

As pointed out by Clarke \textit{et al.}~\cite{Clarke2013}, the measured current contains an ordinary $2\pi$-periodic contribution in parallel with the fractional one, and inelastic processes changing the parafermion occupation can restore $2\pi$ periodicity. We are not aware of any report of a period longer than $2\pi$ in an FQH or FQAH device coupled
to a superconductor. Missing odd Shapiro steps, interpreted as a $4\pi$
periodicity, have been reported in several other systems (see
Ref.~\cite{Dartiailh2021} and references therein).

\subsection{Shapiro steps}
\label{sec:shapiro}

Under microwave irradiation at frequency $f$, voltage steps appear at
$V_n=n\,hf/q^*$, where $q^*=2e/k$ for a junction whose energy has period $2\pi k$
in $\varphi$. The $6\pi$ period corresponds to $k=3$, i.e., $q^*=2e/3$, and the
smaller charge gives the larger spacing:
\begin{align}
q^*=2e:\quad & V_1=hf/2e=20.7\,\mu\text{V},\\
q^*=2e/3:\quad & V_1=3hf/2e=62.0\,\mu\text{V},
\end{align}
at $f=10$~GHz. Because the steps at $3n\,hf/2e$ coincide with every third conventional step, the signature in the presence of the parallel conventional current is the suppression of the steps $V=n\,hf/2e$ with $n$ not divisible by 3, the analog of the missing odd steps of a $4\pi$ junction. The drive must also be slow on the scale of the gap of the trench, $hf\ll E_g$. At $10$~GHz, $hf=41~\mu$eV, which exceeds the smaller estimate of Sec.~\ref{sec:energy}, so lower frequencies may be needed; at $1$~GHz, $V_1=6.2~\mu$V for $q^*=2e/3$.

Anomalous and missing Shapiro steps also occur in non-topological junctions.
Missing odd steps have been observed in trivial InAs junctions and traced to
Landau--Zener transitions in high-transparency channels~\cite{Dartiailh2021}, and
fractional Shapiro steps can appear without a fractional Josephson
effect~\cite{Galaktionov2021}. The step spacing by itself cannot establish
parafermions.

\subsection{Tunneling conductance}

For perfect Andreev conversion at a parafermion domain wall, the zero-bias
conductance is quantized at~\cite{AliceaFendley2016}
\begin{equation}
G=\frac{2\nu e^2}{h}=\frac{4e^2}{3h}\quad(\nu=2/3),
\end{equation}
compared with $2e^2/h$ for Majorana modes. More information is contained in the structure of the peak. Adding a second domain wall splits the zero-bias peak into $g=(\sqrt3)^2=3$ resonances at finite voltage~\cite{AliceaFendley2016}, which reflects the quantum dimension of the walls. Trivial multilevel structures can also produce several resonances, so the count is suggestive rather than conclusive on its own. Since the protected count is 3 in either limit of Sec.~\ref{sec:zeromodes}, the resonance count tests the $\mathbb{Z}_3$ structure without distinguishing the limits. Unprotected states bound to walls where the neutral channel changes would
show up as additional, non-universal resonances. Interferometric probes of the
fusion channels have also been proposed recently~\cite{CaoSohal2026}.

\subsection{Thermal transport}

The thermal Hall conductance measures the net chiral central charge, which vanishes for the $\nu=2/3$ edge, since $K=\mathrm{diag}(1,-3)$ has one downstream and one upstream mode, and gapping the trench does not change it. The parafermion sector leaves no signature in $\kappa_{xy}$. At the self-dual point the trench itself is critical and non-chiral, with $c=4/5$ in the $\mathbb{Z}_3$ parafermion sector ($c=9/5=1+4/5$ for the full critical theory with a fluctuating phase~\cite{Bollmann2026}), and in an ideal ballistic geometry its thermal conductance along the trench would be
\begin{equation}
\kappa=c\,\frac{\pi^2k_B^2T}{3h},
\end{equation}
with Planck's constant $h$, not $\hbar$. For $T=100$~mK and $c=4/5$, $\kappa\simeq7.6\times10^{-14}$~W/K. This requires tuning to criticality, and we do not expect thermal transport to be a practical probe for a device of this size, since it would require a suspended sample, sub-millikelvin thermometry and subtraction of the phonon background.

\section{Intrinsic non-Abelian FCIs}
\label{sec:alternative}

Neither the neutral gap nor the pairing symmetry of Sec.~\ref{sec:fails} is an
issue when the non-Abelian order is intrinsic. Liu, Perea-Causin and
Bergholtz~\cite{Liu2025} give exact-diagonalization evidence (low-energy quantum
numbers, spectral flow, many-body Chern numbers and entanglement spectra) for the
$\mathbb{Z}_3$ Read--Rezayi phase, and hence for Fibonacci anyons, at $\nu=3/5$,
with weaker evidence at $2/5$, in moir\'e minibands with tunable quantum geometry.
This route requires no superconductor, no domain walls and no gapping of a neutral
mode. Fibonacci order has also been proposed in Abelian bilayer quantum Hall states
driven by interlayer tunneling, again without a
superconductor~\cite{VaeziBarkeshli2014}.

The intrinsic route requires reaching $\nu=3/5$ or $2/5$ in a band with nearly
ideal geometry, and as far as we know it has not been observed. Nearly ideal means
close to saturating the inequality
\begin{equation}
\operatorname{tr}g(\bm k)\ \geq\ |B(\bm k)|
\end{equation}
between the Fubini--Study metric and the Berry curvature, which holds pointwise
because the quantum geometric tensor is positive
semidefinite~\cite{Roy2014,Parameswaran2013}. The quantity
$T(\bm k)=\operatorname{tr}g-|B|$ is thus non-negative, and FCIs are favored as
$T\to0$. Jackson, M\"oller and Roy~\cite{Jackson2015} find that the Berry curvature fluctuations and the Brillouin-zone average of $T(\bm k)$ each correlate with FCI stability.

\section{Conclusion}

We have studied the parafermion construction on a trench between two
spin-polarized $\nu=2/3$ edges, the case relevant to FQAH states. Several device
ingredients carry over from the FQH setting: high interface transparencies have
been reached, no fine-tuning is needed, and gate-defined geometries in a single
crystal avoid a heterointerface. An induced pairing gap on a fractional edge has,
as far as we know, not yet been demonstrated. Compared with the unpolarized $(1,1,2)$ state, the essential difference is that equal-spin pairing and electron tunneling are compatible with two different, mutually incompatible neutral channels, whereas singlet pairing and tunneling share one. The two states have the same topological order, and the $(1,1,2)$ singlet pair exists on the polarized trench as a higher-order dressed operator; spin polarization changes which operators are microscopically cheap. Each process generates its own neutral term at second order,
marginal when charge and neutral modes decouple on each edge, and the inter-edge
Coulomb interaction produces both terms at first order. In a clean symmetric trench only the pairing side conserves momentum, which favors a uniform neutral-mode
superconductivity channel. Because the two channels condense the same neutral anyon
up to a local operator, the walls carry $\mathbb{Z}_3$ zero modes, with 3 protected
states per pair, whether the neutral channel is uniform or follows the local
process, and fixing the fermion parity does not enlarge this count. The neutral
channel determines only which charge term enters through a dressed operator and
whether the walls bind additional unprotected states. Once the neutral sector is gapped, both charge terms are
relevant for decoupled edges and remain so for an inter-edge Coulomb coupling below
$7/25$ of the intra-edge coupling. Before that, pairing is at best marginal for repulsive inter-edge coupling unless an attractive neutral coupling favors its channel, and in general the neutral gap has to form first, which is the main open problem of the construction. The superconductor must supply even-frequency,
equal-spin, intravalley pairing in the valley of the FQAH state, which an $s$-wave singlet superconductor with only Ising spin--orbit coupling does not provide at leading order and the chiral triplet states proposed for the superconductor next to the FQAH state
do if they are polarized in the same valley.

The standard experimental signatures are less selective than is often assumed. The
long-period Josephson effect is mixed with conventional contributions and destroyed
by inelastic processes that change the parafermion occupation, and anomalous
Shapiro steps occur in trivial junctions. For $\nu=2/3$ the whole $\mathbb{Z}_3$ label is carried by a quasiparticle of the FQAH state, so neither stray electrons nor any other process confined to the junction changes it, and the fractional Josephson period is $6\pi$ with or without fixed fermion parity. Only processes that reach the far ends of the superconducting regions, quasiparticles entering from the bulk and inelastic transitions can change it, and the tunneling segments must not be FCI. The $12\pi$ period of Laughlin trenches at fixed fermion parity is not expected for $\nu=2/3$. The splitting of the zero-bias conductance peak into three resonances is a more selective test, since it reflects the number of protected states per pair of walls, although it is not conclusive on its own.

\begin{acknowledgments}
\emph{AI disclosure.} The author used Claude Opus 5.5 (Anthropic) for
literature search and checking calculations, and takes full responsibility
for the content.
\end{acknowledgments}

\emph{Data availability.} The scripts for the lattice, scaling-dimension and
junction calculations are provided as ancillary files.

\end{document}